\documentclass[preprints,article,accept,moreauthors,pdftex]{myarxiv} 

\hreflink{https://doi.org/} 

\usepackage{physics}
\usepackage{siunitx}
\newcommand{\sigmaop}{\vec{\boldsymbol{\sigma}}}
\newcommand{\posop}{\vec{\boldsymbol{R}}}
\newcommand{\momop}{\vec{\boldsymbol{P}}}

\Title{Quantum Dynamical Simulation of a Transversal %
       Stern--Gerlach Interferometer} 
\Author{Miko\l{}aj M. Paraniak$^{1,}$ %
        and Berthold-Georg Englert$^{1,2,3}$}

\address{%
$^{1}$ Centre  for  Quantum  Technologies,  Singapore\\
$^{2}$ Department of Physics, National University of Singapore, %
             Singapore\\
$^{3}$ MajuLab, CNRS-UCA-SU-NUS-NTU International Joint Research Unit, %
            Singapore}

\abstract{Originally conceived as a gedankenexperiment, an apparatus
  consisting of two Stern--Gerlach apparatuses joined in an inverted manner
  touched on the fundamental question of the reversibility of evolution in
  quantum mechanics.
  Theoretical analysis showed that uniting the two partial beams requires an
  extreme level of experimental control, making the proposal in its original
  form unrealizable in practice.
  In this work we revisit the above question in a numerical study concerning
  the possibility of partial-beam recombination in a spin-coherent manner.
  Using the Suzuki--Trotter numerical method of wave propagation and a
  configurable, approximation-free magnetic field, a simulation of a
  transversal Stern--Gerlach interferometer under ideal conditions is
  performed.
  The result confirms what has long been hinted at by theoretical analyses:
  the transversal Stern--Gerlach interferometer quantum dynamics is
  fundamentally irreversible even when perfect control of the associated
  magnetic fields and beams is assumed.}  



\makeatletter
\renewcommand{\@evenfoot}{}%
\renewcommand{\@oddfoot}{}%
\renewcommand{\@oddhead}{\normalfont\small %
  \underline{\makebox[\textwidth][c]{M.M.~Paraniak \& B.-G.~Englert\hfill %
  {\itshape Quantum Dynamical Simulation of a Transversal %
    Stern--Gerlach Interferometer}\hfill\thepage}}}%
\renewcommand{\@evenhead}{\@oddhead}%
\makeatletter

\begin{document}

\section{Introduction}

\begin{figure}[t]
\centering
\includegraphics[viewport=140 615 455 760,clip=,scale=1.0]{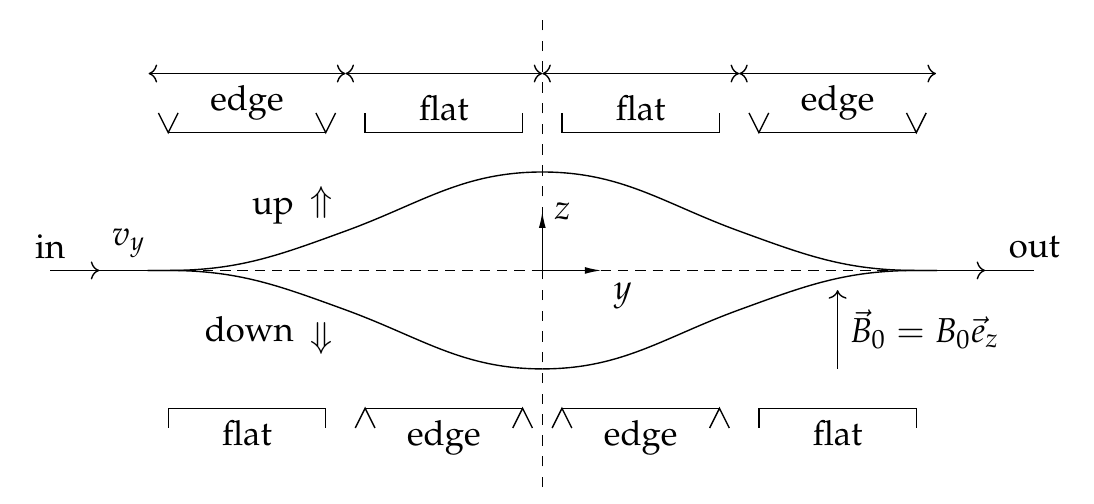}
\caption{\label{fig:sgi-apparatus}%
  A side view of a transversal Stern--Gerlach interferometer as
  envisioned by Bohm (with the modification noted in footnote \ref{fn:Bohm}).
  Four Stern--Gerlach magnets are used to generate a magnetic field $\vec{B}$
  that first splits and then recombines the atomic beam of silver atoms,
  which are moving with velocity $v_y$ through the apparatus.
  The spin quantization axis is determined by the homogenous field $\vec{B}_0$
  oriented along the Z axis.}
\end{figure}

The experiment by Stern and Gerlach performed in 1921 (a centennial this
year)~\cite{stern_weg_1921, gerlach_experimentelle_1922} proved to be of
fundamental importance for the development of quantum mechanics.
Far from being a thing of the past, the original design of the Stern--Gerlach
apparatus (SGA) has continued to inspire new questions
\cite{bloom_transverse_1962, franca_possible_1992, gorceix_dispersive_1994,
  de_oliveira_dissipative_2006, mcgregor_transverse_2011,
  hatifi_revealing_2020, impens_shortcut_2017, mathevet_new_2001,
  xu_phase-dependent_2010, de_carvalho_toward_2015}, and experimental
techniques \cite{robert_stern-gerlach_1995, rubin_atom_2004,
  perales_ultra_2007, amit_t_2019, margalit_realization_2018}.
Professor David Bohm in his well-known textbook on quantum theory discusses
a device consisting of two SGAs (see fig.~\ref{fig:sgi-apparatus}) joined in an
inverted manner so that  
\begin{quote}
    If the uniform magnetic fields (\dots) are set up in exactly the right
    way, and if the second inhomogeneous field is an exact duplicate of the
    first one, the two wave packets can be brought together into a single
    coherent packet.
    Although the precision required to achieve this result would be fantastic,
    it is, in principle, attainable.\\
    \textit{---Section 22.11 in \cite{bohm_1951}}%
   \footnote{\label{fn:Bohm}%
   In Bohm's proposal the atom is deflected in regions of
   ``uniform magnetic fields'' --- but there are no such Lorentz-type forces
   acting on electrically neutral atoms. All subsequent works, including
   ours, consider magnetic gradients to achieve the restoring deflection, with
   three more SGAs completing the interferometer sketched in
   fig.~\ref{fig:sgi-apparatus}. 
   There is also Wigner's scheme, in which the inhomogeneous magnetic field
   from an electric current recombines the beams that emerge from the SGA
   \cite{Wigner:AJP63}; it has not been used for a quantitative model.}
\end{quote}
The analysis of this possibility gave rise to what is known as the
\textit{Humpty-Dumpty problem}
\cite{englert_is_1988,schwinger_is_1988,scully_spin_1989} of coherent
recombination of spatially separated partial atomic beams (i.e. the spin-up
and spin-down components of the spinor wave function) coming out of an SGA, as
well as to a plethora of experimental techniques within the domain of
matter-wave interferometry, all of which use different approaches to work
around the fantastic precision required to realize the original proposal by
Bohm.
Recent advances in atomic chip manufacturing allow the experimenters to
realize a Stern--Gerlach-inspired interferometer
\cite{miniatura_longitudinal_1991, miniatura_longitudinal_1992,
  baudon_atomic_1999, machluf_coherent_2013, amit_t_2019}, whereas the
accuracy required for realizing a transversal Stern--Gerlach interferometer
(SGI) as originally imagined remains fantastic. 

In this work we revisit the question of time-reversibility of a transversal
SGI.
Past theoretical work focused on the impossibility of perfect control of
magnetic fields and beams, indicating a quick loss of spin coherence if such
control could not be maintained \cite{englert_is_1988, schwinger_is_1988}.
Yet, what was also suggested by these analyses is that the very dynamics of a
beam-splitting Stern--Gerlach apparatus might not be reversible as imagined,
even when perfect control over involved magnetic fields and beams is assumed
\cite{englert_time_1997}.
To fill in the missing piece in the quantum Humpty-Dumpty riddle, we have
performed an accurate three-dimensional wave-propagation simulation of a
transversal SGI in order to capture accurately all quantum mechanical effects
that arise once a full, ideal magnetic field is accounted for.
The results of our simulation confirm the intuition that the quantum dynamics
of a Stern--Gerlach apparatus is not reversed by the application of inverse
magnetic fields, even if one has perfect control over the experimental
apparatus.

\section{Modeling the Transversal SGI}
\begin{figure}[t]
  \centering
  \includegraphics[viewport=150 662 445 760,clip=,scale=1.0]{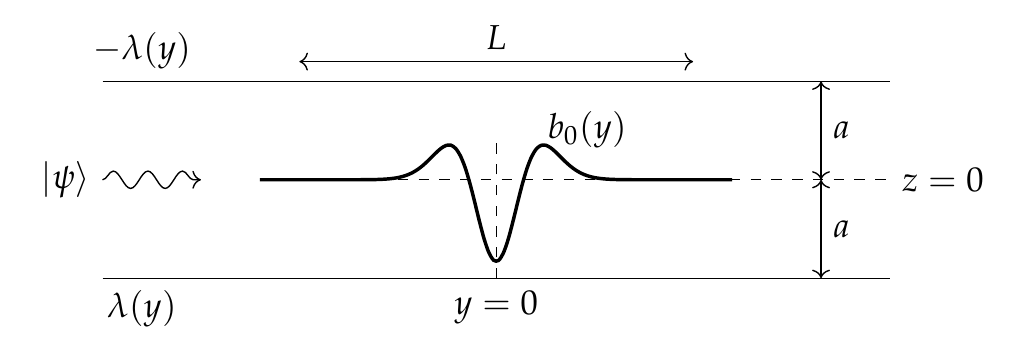}
\caption{\label{fig:sgi-magnetic-field}%
  Magnetic field model used in our simulation.
  In the Y,Z plane we have a fictitious magnetic line charge $\lambda(y)$ at
  $z=-a$ and the opposite charge $-\lambda(y)$ at $z=a$.
  For points on the Y axis, the resulting magnetic field is in the Z direction,
  $\vec{B}=b_0(y)\vec{e}_z$.
  The model is defined by specifying $b_0(y)$;
  see section~\ref{section:magnetic-field}.}
\end{figure}

A beam of silver atoms prepared in a pure spin state directed along the X
axis, $\ket{\psi} = \psi(x,y,z)(\ket{\uparrow} + \ket{\downarrow})/\sqrt{2}$,
with the initial wave function $\psi(x,y,z)$ chosen to be a gaussian wave
packet of width $\delta$,   
\begin{equation}
\begin{aligned} \label{eq:psi-gaussian}
  &\psi(x,y,z) = \frac{1}{\pi^{3/4} \delta^{3/2}}
  \exp(\mathrm{i}ky)
  \exp\Bigl(-\bigl(x^2 + (y-y_0)^2 + z^2\bigr)/\bigl(2\delta^2\bigr)\Bigr),
  \\
  &v_y = \frac{\hbar k}{m}.
\end{aligned}
\end{equation}
The atoms enter the apparatus from the left at time $t = 0$,
at $x=0$, $y=y_0 = -L/2$, $z=0$, travel in the Y direction with velocity
$v_y$, and exit at $y=L/2$, with $y=\pm L/2$ sufficiently far from the SGI
center at $x=y=z=0$ to be outside the magnetic fringing fields.
Inside the apparatus, the magnetic field $\vec{B}$ is generated by two
symmetrically placed linear ``magnetic charge''%
\footnote{By exploiting the analogy between Maxwell's equations for a static
  electric field and a static magnetic field we model the magnetic field
  through a scalar potential as though generated by a charge distribution,
  with the details given in section \ref{section:magnetic-field}.
  It is, of course, also possible to model the magnetic field with adjustable
  electric currents (see, e.g., \cite{Ruiqi-project}) but the
  magnetic-line-charge model is simpler to implement.
  While a parameterization of the magnetic field by an arrangement of a few
  magnetic point charges shares this simplicity \cite{TzyhHaur-project},
  it is less flexible.}
distributions of opposite charge at the height $a$ from the beam's initial
position, according to the arrangement depicted in
fig.~\ref{fig:sgi-magnetic-field}.
The resulting magnetic field is symmetric across the XY plane and the
nonlinear dependence on the $z$ coordinate of the magnetic field due to a
single charge line, which ordinarily would result in slightly different
trajectories for the spin-up and spin-down components, no longer poses an
obstacle for spatial beam recombination.
Furthermore, to fix the spin quantization axis along the Z axis, a bias field
$\vec{B}_0 = B_0 \vec{e}_z$ in the Z direction (here $\vec{e}_z$ is the unit
vector in the Z direction, and $B_0$ is the magnitude of the bias field) is
applied throughout the apparatus.
The magnitude of this field is chosen to be much greater than the magnitudes
of the $B_x$, $B_y$ components in the apparatus to suppress the effects of
Larmor precession around axes other than Z.

The Larmor spin precession angle that the particle accumulates as it travels
through the apparatus results in the final spin of the atom not necessarily
pointing in the X direction despite our best efforts, even if the beams are 
recombined coherently.
It is therefore inappropriate to check whether the final spin state is equal to
the initial one. Instead we measure the degree of preserved coherence, $C(t)$,
by the purity of the statistical operator $M(t)$ for the spin degree of freedom,
\begin{equation}
\begin{aligned}
    \label{eq:spin-coherence}
    M(t) &= \frac{1}{2}\left(1
      + \left\langle{\sigmaop}\right\rangle_{\psi(t)}\cdot\sigmaop\right), \\
    C(t) &= 2\,\mathrm{tr}\left(MM^\dagger\right) - 1
    = \left\langle{\sigmaop}\right\rangle ^2,
\end{aligned}
\end{equation}
where $\sigmaop$ is the vector of Pauli spin matrices.
The spin coherence $C(t)$ ranges between $0.0$ and $1.0$, between a completely
mixed spin state (coherence lost) and a pure spin state (coherence preserved),
with intermediate values indicating some degree of coherence loss. 
After time $T$, when the particle has traversed the length of the apparatus
$L$ and reached the position $y = L/2$, it emerges at the right and its spin
coherence is measured.
It is our goal that the final value $C(T)$ of the spin coherence is as close
to its maximal value $1.0$ as possible; in a perfect experiment we shall have
$C(T) = 1.0$. 

During the evolution the peak separation between the partial beams reaches
$\Delta Z(T/2)$ (henceforth denoted by $\Delta Z$).
One has to ensure that $\Delta Z$ is large compared with the width $\delta$ of
the atom wave packet to say that the partial beams have truly separated.
If despite partial beam separation in the middle of the apparatus the spin
coherence has been preserved, then the evolution in the left half of the
apparatus has been reversed, or in the playful language of
\cite{englert_is_1988}, Humpty-Dumpty, which is really an egg, has been put
back again after his great fall.  
After this overview of the experiment we move on to describing the methods
used in the simulation.

\subsection{Quantum Dynamics of the Apparatus}
Following the usual quantum treatment of an SGA, we have the following
Hamiltonian describing the spin-1/2 silver atom comprising the beam:
\begin{equation}\label{eq:Hamiltonian}
    \mathcal{H} = \frac{\momop^2}{2m} - \mu \sigmaop\cdot\vec{B}(\posop)
\end{equation}
with the momentum operator $\vec{\boldsymbol{P}} = (\boldsymbol{P}_x,
\boldsymbol{P}_y, \boldsymbol{P}_z)$,
the position operator $\vec{\boldsymbol{R}} = (\boldsymbol{R}_x,
\boldsymbol{R}_y, \boldsymbol{R}_z)$, the magnetic moment $\mu$,  
the vector of Pauli spin matrices $\sigmaop = (\boldsymbol{\sigma}_x,
\boldsymbol{\sigma}_y, \boldsymbol{\sigma}_z)$, and the position-dependent
magnetic field $\vec{B}(\vec{\boldsymbol{R}})$ in the apparatus.  
The numerical values for $m$ and $\mu$ that are used in the simulation, and
other simulation parameters are included in the appendix.  

To solve the equations of motion with an accuracy high enough to model
coherent beam recombination we employ a fourth-order Suzuki--Trotter operator
splitting method \cite{suzuki_generalized_1976}.
In this scheme, given any Hamiltonian with a potential-energy term
$V(\posop)$, the time propagator
$U = e^{-\mathrm{i}t/\hbar\,(\momop^2/(2m) + V(\posop))}$, which advances the
wave function by a time step $t$, is split into several exponential terms,
each involving only the momentum operator or the potential operator with
coefficients specially chosen, such that the whole formula is accurate to a
certain order $k$, 
\begin{equation}\label{eq:gen-ST}
  U_N = \prod^{\left\lceil N/2 \right\rceil}_{i=1}
  \mathrm{e}^{-\frac{\mathrm{i}t}{\hbar} \alpha_i V(\posop)}
  \mathrm{e}^{-\frac{\mathrm{i}t}{\hbar} \beta_i \frac{\momop^2}{2m}}
  = \mathrm{e}^{-tA_1}\mathrm{e}^{-tB_1} \mathrm{e}^{-tA_2}
  \mathrm{e}^{-tB_2} \cdots,
\end{equation}
where $A_i$, $B_i$ correspond respectively to potential and momentum operator
terms.  
While approximations $U_N$ for increasing $N$ yield higher-order formulas and
can in principle achieve any desired level of accuracy, the computational cost
goes up considerably with increasing $N$, as between each application of
operators either in momentum or in position space one has to perform a Fourier
transformation on the wave function.
For our purpose we have chosen a fourth-order Suzuki--Trotter method of a
special form that involves a gradient term of the potential and five operator
terms in total, compared with 11 terms in the standard Suzuki--Trotter formula
needed to achieve the same order \cite{hatano_finding_2005,
  chin_symplectic_1997,hue_fourth-order_2020}.  
The time propagator in this ``$7\star$ approximation'' is%
\footnote{The notation $U_{7\star}$ is a reminder that this is a limiting case
  of a gradient-free seven-factor approximation $U_7$ as in
  eq.~(\ref{eq:gen-ST}); see \cite{ChauTT+3:2018} for the technical details.}
\begin{equation}
    \begin{split}
      U_{7\star} =
      &\exp(-\frac{\mathrm{i}t}{6\hbar} V(\posop))
      \exp(-\frac{\mathrm{i}t}{4\hbar m} \momop^2) \\
      &\times \exp(-\frac{\mathrm{i}t}{\hbar}
      \left(\frac{2}{3} V(\posop) - \frac{1}{72m}
        \left[t \grad{V}(\posop)\right]^2\right)) \\
      &\times \exp(-\frac{\mathrm{i}t}{4\hbar m} \momop^2)
      \exp(-\frac{\mathrm{i}t}{6\hbar} V(\posop)), 
    \end{split}
\end{equation}
and to propagate the wave function the above five unitary operators are
applied in their respective space.
For instance, the application of the first two terms from the right
\begin{equation*}
  \exp(-\frac{\mathrm{i}t}{4\hbar m} \momop^2)
  \exp(-\frac{\mathrm{i}t}{6\hbar} V(\posop)) 
\end{equation*}
on $\psi(\vec{R}, t_0)$ as part of the algorithm requires two
Fourier transformations $\mathcal{F}$, 
\begin{equation}
  \psi(\vec{R}, t_0 + t)
  = \cdots\mathcal{F}^{-1}\left\{\exp(-\frac{\mathrm{i}t}{4\hbar m} \momop^2)
    \mathcal{F}\left\{\exp(-\frac{\mathrm{i}t}{6\hbar} V(\posop))
      \psi(\vec{R}, t_0)\right\}\right\}.
\end{equation}
A single application of the $U_{7\star}$ propagator requires the execution of
four Fourier transformations for a spinless wave function.  
Time propagation of a spin-1/2 particle will require eight Fourier
transformations per time step, an operation that contributes heavily to the computational cost. 

Once this split-operator machinery is in place we can simulate the quantum
evolution of any system described by a reasonable Hamiltonian with a
position-dependent potential, including the quantum dynamics of an SGI.
However, when attempting to do so according to the method presented above, one
is quickly faced with rising computational space and time resources required
to perform the computation to a sufficient degree of accuracy.
The storage cost of a wave function across the region of the apparatus is
prohibitive, as is the time required to perform repeated Fourier
transformations in each time step.  

To make the calculation feasible, we transform the Hamiltonian into the
co-moving frame of the atomic beam traveling with velocity $v_y$ along the Y
axis of the apparatus.  
This is done at the cost of making the potential term in the Hamiltonian
time-dependent, 
\begin{equation}
    \label{eq:sgi-time-hamiltonian}
    \mathcal{H} = \frac{\momop^2}{2m}
    - \mu \sigmaop\cdot\vec{B}(\boldsymbol{X}, y_0 + v_y\,t - \boldsymbol{Y},
    \boldsymbol{Z}),
\end{equation}
and thus requires a re-evaluation of the split-operator method which,
presented in its most basic form, is applicable only to Hamiltonians not
explicitly dependent on time.
The extension of the split-operator method to time-dependent non-commuting
Hamiltonians is treated in full generality in \cite{suzuki_general_1993}.  
Here we just note the elegant result: in the time-dependent split-operator
method $U_{7\star T}$ the reference time has to be advanced cumulatively at
each encountered $\exp(-\frac{\mathrm{i}t\beta}{\hbar} \frac{\momop^2}{2m})$
term, $T^\prime = T + \beta\,t$, so that subsequently applied potential
operator terms use the newly advanced time $T^\prime$, until the next
application of a position propagator and so on.
The resulting formula for the time-dependent propagator is then guaranteed to
be accurate to the same order as the original formula.
Using this simple prescription we readily obtain the time-dependent propagator
$U_{7\star T}$, for the evolution from time $T$ to time $T + t$, 
\begin{equation}
    \begin{split}
      U_{7\star T} = &\exp(-\frac{\mathrm{i}t}{6\hbar} V(\posop, T + t))
      \exp(-\frac{\mathrm{i}t}{4\hbar m} \momop^2)\\
      &\times \exp(-\frac{\mathrm{i}t}{\hbar}
      \left(\frac{2}{3} V(\posop, T + t/2) - \frac{t^2}{72m}
        \left[\grad{V}(\posop, T + t/2)\right]^2\right))\\
      &\times \exp(-\frac{\mathrm{i}t}{4\hbar m} \momop^2)
      \exp(-\frac{\mathrm{i}t}{6\hbar} V(\posop, T)).
    \end{split}
\end{equation}
In this way one has to keep track only of the wave function in the region
required to model the beam deflection in the Z axis, which is very small
compared with the total extent of the apparatus, as the movement along the Y
direction has been taken care of by transforming the Hamiltonian into the
co-moving frame.
The outstanding challenge is an efficient and accurate computation of the
magnetic field $\vec{B}(x,y,z)$, which in each time step has to be evaluated
across the domain of the wave function, now in the co-moving frame of the
beam.

\subsection{Magnetic Field in the Apparatus}
\label{section:magnetic-field}
To our best knowledge all previous quantum treatments of Stern--Gerlach-like
experiments employed a magnetic field that either did not satisfy the Maxwell
equations through the neglect of the X and Y field components and linear
approximations \cite{englert_is_1988, schwinger_is_1988, japha_general_2019},
or used a simple model of a magnetic field that neglects the fringing fields
at the entrance and exit of the apparatus \cite{manoukian_quantum_2003,
  hsu_stern-gerlach_2011}. 
For the purpose of our work it is crucial that the employed field model is
free from both of these approximations, since it is clear that either of these
simplifications will result in an apparatus that will seemingly have no
problem with reuniting the partial beams in a coherent manner, as long as the
beams and fields are accurately controlled. 

In the traditional design, at least four Stern--Gerlach magnets have to be
employed.
As the particle goes through the apparatus, the partial beams corresponding to
the spin-up and spin-down components are first imparted a momentum which leads
to their spatial separations as seen in the schematic trajectories in
fig.~\ref{fig:sgi-apparatus}.
Subsequently, the evolution both in momentum and position has to be
reversed. The second and third magnets first decelerate the particle and then
impart momentum equal in magnitude to the momentum gained in the first stage,
and finally the last magnet decelerates the particle, whereupon, if the
conditions were ideal, we should find the partial beams reunited, and the
particle in a pure spin state. 

In our scheme, we exploit the correspondence between the Maxwell equations for
the electric field and the static magnetic field,
\begin{equation}
    \begin{aligned}
        &\curl \vec{E} = 0, \quad &\div\vec{E} = 0, \\
        &\curl \vec{B} = 0, \quad &\div\vec{B} = 0,
    \end{aligned}
\end{equation}
and introduce a fictitious magnetic ``charge'' potential $\phi$, which obeys
the Laplace equation 
\begin{equation}
    \label{eq:phi-laplacian}
    \grad^2 \phi = 0,
\end{equation}
and results in a physical magnetic field
\begin{equation}
    \vec{B} = -\grad{\phi},
\end{equation}
in full analogy to a static electric field.
In this manner we obtain a configurable model of a magnetic field free from
approximations.
This is in contrast to other approaches to the Humpty-Dumpty problem, which
tend to turn the problem of reversing the evolution into a technical
difficulty depending solely upon the experimenter's skill, thereby possibly
concealing difficulties of a more fundamental nature, while our aim is
precisely to uncover those difficulties that arise once all simplifications
are put aside.  

The scalar potential for the pair of magnetic line charges (see
fig.~\ref{fig:sgi-magnetic-field}) is
\begin{align}
  \label{eq:eq:phi-lambda}
  \phi(\vec{r})&=\int_{-\infty}^{\infty}\mathrm{d}y'\,\lambda(y')
  \Bigl(\bigl[(y-y')^2+s_+^2\bigr]^{-1/2}
        -\bigl[(y-y')^2+s_-^2\bigr]^{-1/2}\Bigr)
                 \nonumber\\
   &=\int_{-\infty}^{\infty}\frac{\mathrm{d}k}{\pi}\, \mathrm{e}^{-\mathrm{i}ky}
                 \bigl[K_0\bigl(|k|s_+\bigr)-K_0\bigl(|k|s_-\bigr)\bigr]
                  \int_{-\infty}^{\infty}\mathrm{d}y'\,
     \mathrm{e}^{\mathrm{i}ky'}\lambda(y')\,,
\end{align}
where
\begin{equation}
  \label{eq:s+-}
  s_{\pm}=\sqrt{x^2+(z\pm a)^2}
\end{equation}
are the respective distances from the charged lines,
and $K_0(\ )$ is the zeroth-order modified Bessel function of the second kind.
On the Y axis, we have
\begin{equation}
  \label{eq:grad-phi}
  -\grad\phi\Bigr|_{x=z=0}
  =-\frac{\partial\phi}{\partial z}\Bigr|_{s_{\pm}=a}\vec{e}_z=2b_0(y)\vec{e}_z
\end{equation}
with $b_0(y)$ and $\lambda(y)$ related to each other by
\begin{equation}
  \label{eq:b0}
    \int_{-\infty}^{\infty}\mathrm{d}y\,\mathrm{e}^{\mathrm{i}ky} \lambda(y)=
  \frac{1}{2|k|\,K_1\bigl(|k|a\bigr)}
  \int_{-\infty}^{\infty}\mathrm{d}y\,\mathrm{e}^{\mathrm{i}ky} b_0(y)\,,
\end{equation}
where $K_1(\ )$ is the first-order modified Bessel function of the second kind.
Upon expressing $\phi(\vec{r})$ in terms of $b_0(y)$,
\begin{equation}
  \label{eq:phi-b0}
  \phi(\vec{r})
  =\int_{-\infty}^{\infty}\frac{\mathrm{d}k}{2\pi}\,
  \mathrm{e}^{-\mathrm{i}ky}
  \frac{K_0\bigl(|k|s_+\bigr)-K_0\bigl(|k|s_-\bigr)}
  {|k|\,K_1\bigl(|k|a\bigr)}
  \int_{-\infty}^{\infty}\mathrm{d}y'\,\mathrm{e}^{\mathrm{i}ky'} b_0(y')\,,
\end{equation}
one confirms that $\phi(\vec{r})$ satisfies the Laplace equation
\eqref{eq:phi-laplacian} for $s_{\pm}>0$ by an exercise in differentiation, for
which the relations \cite{abramowitz_1964} 
\begin{equation}
    K'_0(z) = -K_1(z), \qquad
    K'_1(z) = -\frac{1}{z} K_1(z) - K_0(z)
\end{equation}
are useful.

After including the bias field $B_0\vec{e}_z$ and a strength parameter $f$,
the magnetic field is
\begin{equation}\label{eq:def-f}
    \vec{B} = (-\grad{\phi} + B_0\vec{e}_z)\,f.
\end{equation}
Rather than specifying the line-charge density $\lambda(y)$,
we define the model by the choice for $b_0(y)$, essentially
$\vec{B}=B_z\vec{e}_z$ along the level trajectory of the atom that traverses
the SGI.
This makes it easier to ensure that the magnetic field decreases rapidly
outside the SGA.
Indeed, as we show below, one can model a magnetic field suitable for
coherent beam recombination by an appropriate choice for $b_0(y)$, the
strength parameter $f$, and the separation ${d = 2a}$
between the two line charges.

\subsection{Beam Recombination in the Calibrated Magnetic Field}
The conventional treatment of an SGA assumes that the atom's deflection is
caused by a linear $B_z \propto z$ field, thus making the force $F_z \propto
\frac{\partial B_z}{\partial z}$ position independent, which greatly aids in
subsequent beam recombination.
However, a field like this is unphysical, it does not satisfy the Maxwell
equations, and thus conclusions regarding the reversibility of the SGA's
dynamics based on reasoning which neglects this issue are unconvincing as they
risk missing quantum effects that might have an impact on the spin coherence.  
A physical field will necessarily have a non-linear dependence on the $z$
coordinate and thus requires fine tuning to make the partial beams come back
to zero separation.  

In our scheme one can calibrate the field to result in a desired peak
separation $\Delta Z$ by adjusting the field strength parameter $f$ and the
distance $2a$ between the top and bottom line charges.
For $b_0(y)$ we choose a magnetic field profile composed of two gaussians
with an adjustable width parameterized by $\alpha$ (as shown schematically in
\mbox{fig.~\ref{fig:sgi-magnetic-field}}),
\begin{equation}
    \label{eq:b0}
    b_0(y) = \frac{\alpha}{\sqrt{2\pi}}
    \biggl[\exp\Bigl(-(\alpha y)^2 / 2\Bigr)
    - \sqrt{2} \exp\Bigl(-(\alpha y)^2\Bigr)\biggr]
\end{equation}
The parameter $\alpha$ determines the longitudinal stretch of the apparatus over which the particle is deflected, influencing the overall deflection, as well as the range of the fringing fields --- the wider the gaussian peaks, the slower the decay of the magnetic field off the Y axis. As such, a judicious choice of $\alpha$ is needed, firstly to avoid a gaussian that is too narrow and expensive to resolve numerically, and secondly to avoid a gaussian too wide, which then makes it necessary to model an apparatus sufficiently wide to have the fringing fields negligible at both ends of the apparatus.

\begin{figure}[t]
\centering
 \includegraphics[viewport=170 562 430 760,clip=,scale=1.0]{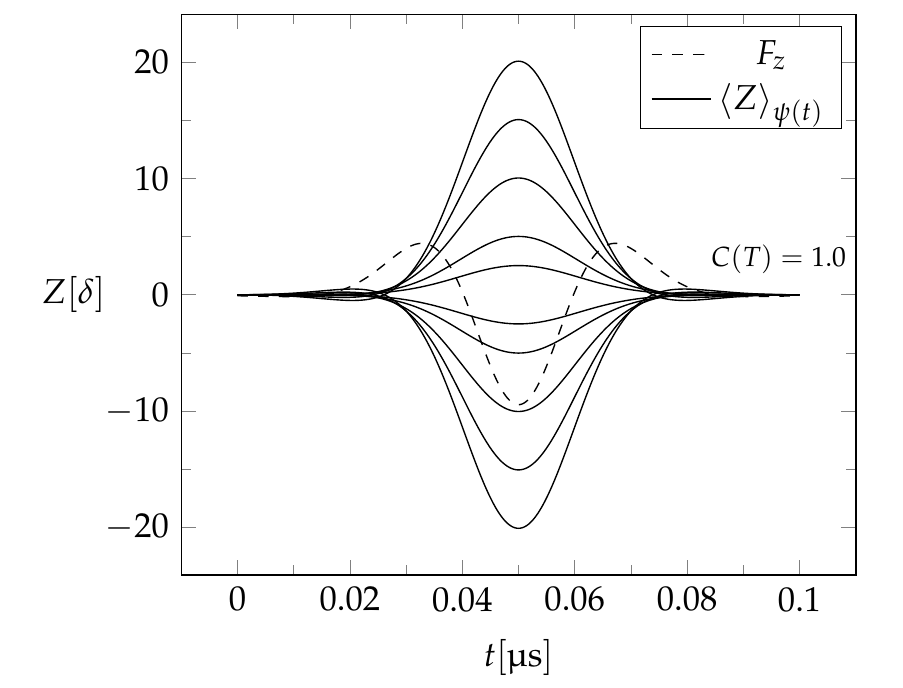}
 \caption{\label{fig:sim-simplified}%
   Trajectories of the partial beams for beam separations
   $\Delta Z = 5, 10, 20, 30, 40 [\delta]$, according to the quantum dynamics
   of the simplified Hamiltonian.
   The spin-up and spin-down partial trajectories for each peak separation
   $\Delta Z$ are symmetric.
   The final coherence is rounded to four decimal places and 
   is $C(T) = 1.0$ regardless of the magnitude of peak separation. 
   The slightly visible ribbon shape of the trajectories near the input and
   output of the apparatus for higher peak separations is a feature of  
   the fringing fields of the employed magnetic field.
   The dashed line is the force $F_z$, in units of $\mu\,b_0(y=0)/a$, acting
   on the partial spin-up beam of the largest separation $\Delta Z =
   40\delta$.} 
\end{figure}

The force in the Z direction resulting from such a profile, as shown in
fig. \ref{fig:sim-simplified}, effectively consists of four regions of unequal
length symmetric accross the Z axis.
These regions are indicated by the direction of the force in each sector:
$+$, $-$, $-$, $+$.
Such an arrangement allows us to mimic the effect of the four-magnet setup of
fig.~\ref{fig:sgi-apparatus} in a simple way, without having 
to introduce a model of the magnetic field for each SG magnet separately.
Note that this $b_0(y)$ is even in $y$ and, therefore, we can replace the
exponential factors in eq.~\eqref{eq:phi-b0} by their real parts. 

The field felt by the particle along its trajectory is approximately $B_z
\approx b_0(y)$, as long as the deflections are much smaller than the length
scale of the field changes.
Then, the vanishing of the integral
\begin{equation}
    \label{eq:b0-integral}
    \int^{\infty}_{-\infty} \mathrm{d}y\, b_0(y) = 0
\end{equation}
makes the Larmor precession around the Z axis approximately equal to 0, with
the spin pointing in the X direction at the output.  
Beam recombination along the Y axis is facilitated by energy conservation ---
changes in potential energy coming chiefly from the changes in $B_z \approx
b_0(y)$ along the trajectory in the left half of the interferometer
(fig.~\ref{fig:sgi-apparatus}) are reversed by the opposite changes in the
right half (since $b_0(-y) = b_0(y)$).
However, the choice $b_0(y)$ dictated by this reasoning is not a sufficient
guarantee of coherent recombination.
The real trajectory is not level and becomes minutely sensitive to the exact
field, non-linear in its $z$ coordinate dependence, in the apparatus.
With the two remaining parameters at hand, the line charge separation $2a$ and
the field strength $f$, we perform a bisection search for these parameters,
constrained by the desired value of the peak separation $\Delta
Z(T/2)\equiv\Delta Z$ and
with the objective of vanishing final spatial separation of the beams, 
\begin{equation}
    \Delta Z (T) \le 10^{-6}\delta.
\end{equation}
This is done in a semiclassical manner: the Z axis quantum and classical
dynamics are in good agreement, the classical simulation being much cheaper to
perform than the quantum counterpart, thus making this optimization scheme
feasible.   
We obtain a range of parameters $a$, $f$ that result in spin-coherent
recombination for different values of the for peak separation,
$\Delta Z = 1\delta, 2\delta, \dots, 40\delta$. 
With this set of optimized parameters $\{a, f\}$ at hand (their values can be
found in the appendix), each choice resulting in a suitable magnetic field and
trajectories with peak separation $\Delta Z_{a,f}$, we can now proceed to
simulating the quantum dynamics of the SGI. 

\enlargethispage{1.0\baselineskip}

\section{Quantum Dynamical Simulation of the SGI}
Using the optimized magnetic fields and the methods discussed in the preceding
sections, we can now perform the wave-propagation simulation of the SGI.
In the first step, we performed a quantum simulation of the beam's
evolution in the optimized fields neglecting the microscopic quantum motions
in the X, Y axes.
The effective Hamiltonian for this case is a simplified one-dimensional version
of the full Hamiltonian of eq.~\eqref{eq:sgi-time-hamiltonian}: 
\begin{equation}
    \label{eq:sgi-simple-hamiltonian}
    \mathcal{H} = \frac{\boldsymbol{P}_z^2}{2m}
    - \mu \boldsymbol{\sigma_z} B_z(0, y_0 + v_y\,t, \boldsymbol{Z}).
\end{equation}
The simulated beam trajectories are presented in fig.~\ref{fig:sim-simplified}.

\begin{figure*}[t]\centering
  \includegraphics[viewport=70 592 520 760,clip=,scale=1.0]{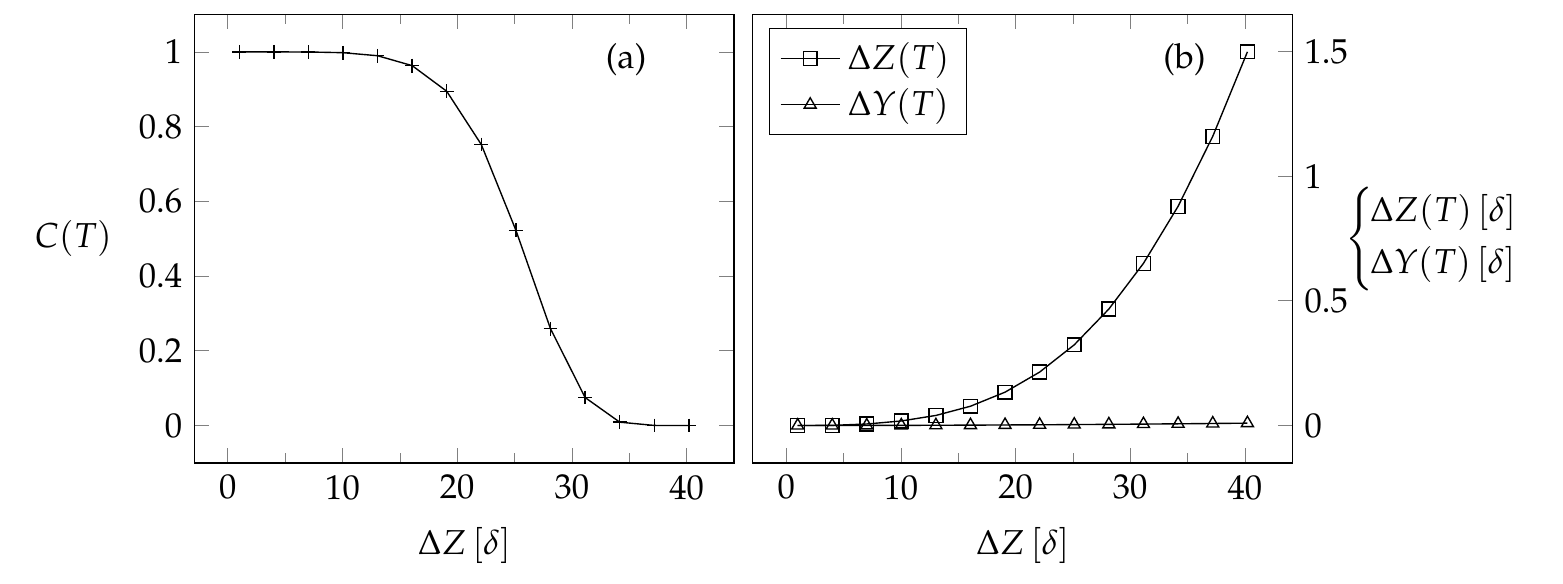}%
\caption{\label{fig:sim-separation-coherence}%
    The final spin coherence $C(T)$ in panel (a) and relative partial beam
    displacements $\Delta Z(T)$, $\Delta Y(T)$ in panel (b) as a function of
    increasing peak separation $\Delta Z$.  
    The magnetic field that results in perfect beam recombination when the
    microscopic quantum movements in the X and Y axis are neglected, is no
    longer sufficient to reverse the full quantum evolution when $\Delta Z$
    becomes appreciable.}     
\end{figure*}

Regardless of the magnitude of the peak separation $\Delta Z = 1\delta,
2\delta,\dots, 40\delta$ the beams are recombined coherently. 
Note that this happens only because the magnetic field, which depends on the
$z$ coordinate in a non-linear fashion (making the force $F_z$ position
dependent), is finely calibrated for each $\Delta Z$ separately.  
The standard recipe of taking the ``reverse'' fields to reverse the evolution
of the beam that underwent the splitting by SG magnets, therefore seems to
work, as long as the fields are calibrated in an ``exactly right'' way.
But is it really true?  
The standard argument assumes that the neglected X and Y movements are
miniscule compared with the large deflection in the Z axis and, therefore,
should not influence the overall analysis.
However, since coherent beam recombination is concerned precisely with the
relative microscopic realignment of the partial spin-up, spin-down beams, it
is necessary to simulate the full dynamics.  

We calculated the quantum dynamics of the original Hamiltonian of
eq.~\eqref{eq:sgi-time-hamiltonian}, again varying the peak separation $\Delta
Z$ and measuring the final spin coherence $C(T)$.
The peak separation-spin coherence curve, together with the final relative
displacements of the partial beams in $y$, $z$ coordinates are presented in
fig.~\ref{fig:sim-separation-coherence}.
Although for very small separations up to $\Delta Z = 10\delta$ the partial
beams are recombined coherently---note that this does not mean that the
evolution was fully reversed: the spreading of the wave packet, though
inconsequential, has not been undone---once the peak separation exceeds this
value, the evolution rapidly becomes irreversible, with $C(T) = 0.0$ indicating
complete spin decoherence by the time separation reaches $\Delta Z =
35\delta$.  
The cause of this loss of spin coherence is the growing microscopic separation
in the Z and Y axes between the up and down spin components as shown in
fig.~\ref{fig:sim-separation-coherence}b, which, although barely exceeding
$\delta$ in Z and negligible in Y, is enough to completely prevent coherent
beam recombination.

\section{Discussion and Conclusion}
An SGA entangles the spin degree of freedom of a spin-1/2 atom with its
orbital degrees of freedom, so that the outcome of a position measurement
informs us about the spin of the atom (``up'' or ``down'' in the traverse
direction that is probed).
The transition from the initial not-entangled state to the final entangled
state is unitary; it is not an irreversible event%
\footnote{Haag emphasized the fundamental role of events \cite{Haag:CMP90} and
  their random realization \cite{Haag:FP13}; see also \cite{Englert:EPJD13}.}
that can be amplified and recorded.
Since unitary processes compose a group, they are invertible in a
\emph{mathematical} sense.
Is the entangling action of a SGA also invertible as a \emph{physical} process
--- by another unitary transition, that is?%
\footnote{Processes that trap the atom and re-prepare it in the initial
  not-entangled state are not unitary and do not count.}

``Yes,'' answered Bohm \cite{bohm_1951} and later Wigner \cite{Wigner:AJP63},
and others parroted their wisdom.
While Bohm acknowledged that the reversal would require a ``fantastic''
precision in controlling the apparatus and Wigner observed that such an
experiment ``would be difficult to perform,'' both took the reversibility for
granted and did not elaborate on the matter.
It is, however, unclear whether elementary unitary quantum processes are
reversible --- in marked contrast to macroscopic processes, which are known to
be irreversible, as witnessed by the folk wisdom of the medieval Humpty-Dumpty
rhyme, the laws of thermodynamics, the lessons of deterministic chaos, the
ubiquitous ageing of living organisms, and other familiar phenomena.

In the particular situation of an SGA, the atom's spin and orbital degrees of
freedom are coupled by the force resulting from the magnetic field gradient,
and only this coupling is available for reversing the action of the SGA in a
unitary fashion.
We cannot undo the action of the unitary evolution operator
$\exp\bigl(-\mathrm{i}tH/\hbar\bigr)$, with $H$ as in
eq.~\eqref{eq:Hamiltonian}, by applying its inverse
$\exp\bigl(-\mathrm{i}t(-H)/\hbar\bigr)$ because $-H$ is not a physical
Hamiltonian --- so much for \emph{mathematical} reversibility.%
\footnote{More about this can be found in \cite{englert_time_1997} and Section
  5.2 in \cite{Englert:EPJD13}.} 
Rather, we must carefully tailor the magnetic field to implement the
disentangling reversal as well as we can.

We recall that Bohm's ``fantastic'' accuracy amounts to controlling the
macroscopic devices with submicroscopic precision \cite{englert_is_1988}, and
if such control is granted, Maxwell's equations prevent us from undoing the
entanglement perfectly \cite{schwinger_is_1988}, even if we neglect the
effects of fringing fields (as is the case in \cite{schwinger_is_1988}).
These theoretical works are here supplemented by a numerical study that fully
accounts for the quantum dynamics with a magnetic field that obeys Maxwell's
equations throughout the volume probed by the atom.

Thereby we assume that there are no uncontrolled imperfections.
This is, of course, an over-idealization that lacks justification (as it
ignores the lesson of \cite{englert_is_1988}).
But even with this stretch we find that the action of the SGA cannot be undone
if the SGA serves its purpose, namely separates the atoms into well
distinguishable partial beams --- $\Delta Z>20\delta$, say.

In conclusion, the accurate wave-propagation results presented here
quantitatively confirm that the microscopic quantum dynamical effects in a
transversal SGI apparatus are enough to quickly destroy spin coherence beyond
recovery. 
The deceptively simple nature of the Stern--Gerlach beam-splitting apparatus
and the seemingly obvious proposal to reverse its evolution by employing
``reverse'' fields might lead one to assume the fundamental reversibility
of its quantum evolution, as long as the apparatus is controlled sufficiently
well.
Such assumptions are, however, unwarranted.

\section*{Acknowledgments}
We thank Jun Hao Hue for valuable discussions.
The efforts by Tzyh Haur Yang and Ruiqi Ding, who conducted closely related
undergraduate projects, are gratefully acknowledged.

\appendixtitles{no} 
\appendixstart
\appendix

\section*{Appendix: Particle properties and initial wave function}
As in the original Stern--Gerlach experiment, we used a beam of silver
atoms to simulate the splitting and recombination.
The values of the various variables are given in table~\ref{tbl:atom-wf}. 
In the lapse of time $T$, the width of the wave function would increase by a
factor of 
\begin{equation}
  \label{eq:spreading}
  \sqrt{1+{\left(\frac{\hbar T}{m\delta^2}\right)}^2} \approx 1.16
\end{equation}
if there were force-free motion, and this amount of spreading is of no
consequence.

\begin{specialtable}[b]\centering
  \caption{\label{tbl:atom-wf}%
    Values used in the simulation: atom properties, apparatus geometry,
    and initial wave function.}\vspace*{-1ex}
  \begin{tabular}{llll}
    &&first&\\[-0.5ex] 
    variable&value&occurrence&description\\ \hline
    $m$ & $107.8682\,\si{Da}$ & in eq.~\eqref{eq:psi-gaussian}
        & mass of a silver atom ($\si{Da}$: dalton)\rule{0pt}{10pt}\\
    $\mu$ & $\frac{1}{2}\mu_{\mathrm{B}}$ & in eq.~\eqref{eq:Hamiltonian}
         & magnetic moment ($\mu_{\mathrm{B}}$: Bohr magneton)\\[1ex]
    $L$ & $25\,\si{\micro \meter}$ & after  eq.~\eqref{eq:psi-gaussian}
        & length of the SGI\\ 
    $T$ & $0.1\,\si{\micro\second}$ & after eq.~\eqref{eq:spin-coherence}
        & time to traverse the SGI\\[1ex]
    $y_0=-L/2$ & $-12.5\,\si{\micro \meter}$ & in eq.~\eqref{eq:psi-gaussian}
               & initial position\\
    $v_y=L/T$ & $ 250\,\si{\meter\,\second^{-1}}$
                  & in eq.~\eqref{eq:psi-gaussian} & initial velocity\\
    $\delta$ &  $0.01\,\si{\micro\meter}$ & in eq.~\eqref{eq:psi-gaussian}
             & width of the initial wave function
  \end{tabular}
\end{specialtable}

\section*{Appendix: Magnetic Field}
\begin{figure}[h]\centering
  \includegraphics[viewport=200 410 500 630,clip=,scale=1.0]{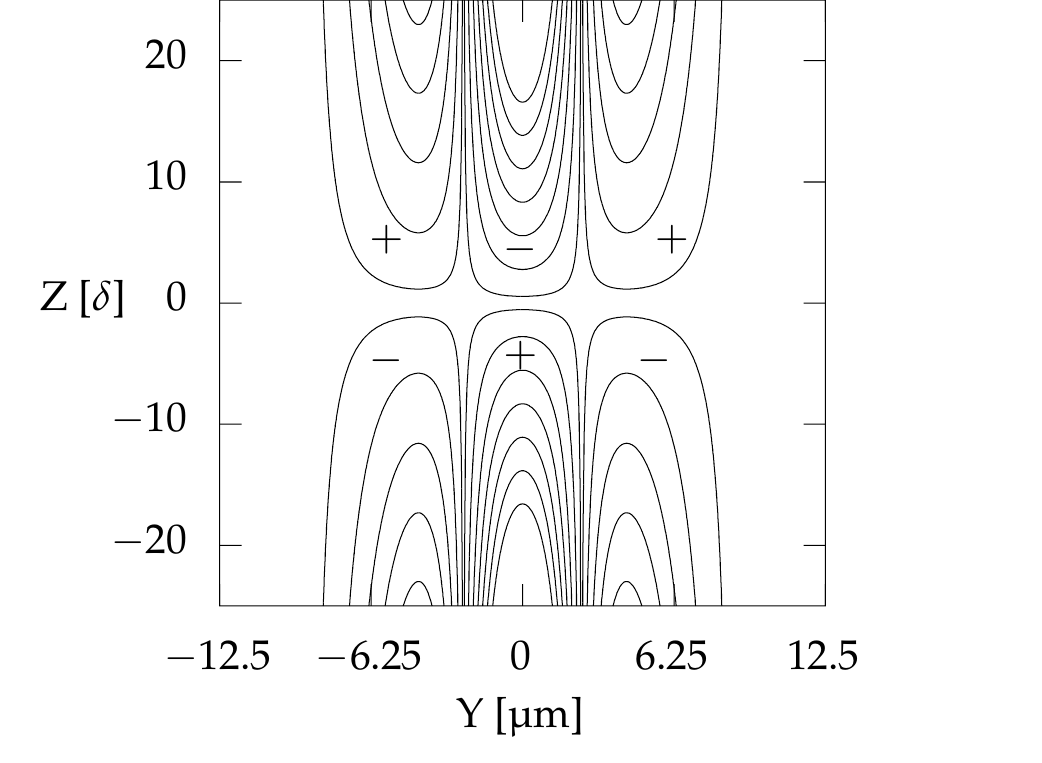}
  \caption{\label{fig:magnetic-field-contour}%
    Lines of constant $B_z$ of the magnetic field calibrated for $\Delta Z =
    40\delta$ shown in the YZ plane.
    The sign of the field in a particular region is indicated by a $+$ ($B_z >
    0$) and $-$ ($B_z < 0$).
    The nested contours increase in field strength as the field gets closer to
    the line charges located at a distance $\pm a$ at the top and bottom,
    according to table \ref{table:optimal-field-parameters}.
    Note that the width of the gaussian wave packet $\delta$ is small compared
    with the overall changes of the field.
    In the middle of the apparatus ($Z = 0$ line) the field vanishes exactly 
    because the line charges of opposite sign are placed symmetrically across
    the XY plane. }
\end{figure}

The parameter $\alpha$ of the magnetic profile function $b_0(y)$ of
eq.~\eqref{eq:b0}, depicted in fig.~\ref{fig:sgi-magnetic-field},
was chosen to be 
\begin{equation}
  \alpha = 0.3\,\si{\micro\meter}^{-1}.
\end{equation}
The example parameters of the optimized fields, rounded to five decimal
places, are displayed in table~\ref{table:optimal-field-parameters}.
In fig.~\ref{fig:magnetic-field-contour} the field lines of the $B_z$
component of the field calibrated for $\Delta Z = 40\delta$ (the largest
simulated peak separation) are shown.
The overall magnitude of the $B_y$ component of the generated field is
actually larger than that the magnitude of $B_z$.
By adjusting $B_0$ we make sure that the resulting $B_z$ field component is
the largest magnetic field component in the apparatus.
For the fields used in the simulations the suitable value of $B_0$ in units of
$b_0(y=0)$, where a.u. denotes atomic units of magnetic field strength, is 
\begin{equation}
  B_0 = 100.0\,\textrm{a.u.} \big/ \bigl|b_0(y=0)\bigr| \approx 2017.
\end{equation}

\begin{specialtable}[t]\centering
  \caption{\label{table:optimal-field-parameters}%
    Optimized magnetic field parameters: beam separation $\Delta Z$
    (cf.\ fig.~\ref{fig:sim-simplified}),
    field-strength parameter $f$ of eq.~\eqref{eq:def-f},
  and line-charge distance $a$ (cf.\ fig.~\ref{fig:sgi-magnetic-field}).}
\begin{tabular}{ccc}
  $\Delta Z [\delta]$ & $f$ &  $a [\si{\micro\meter}]$\\
    \hline
    \phantom{2}5 & 0.00011 & 0.71919\rule{0pt}{12pt} \\
    10 & 0.00030 & 0.04353 \\
    20 & 0.00081 & 1.61404 \\
    30 & 0.00147 & 2.19168 \\
    40 & 0.00226 & 2.85283 \\
\end{tabular}
\end{specialtable}

The evaluation of the magnetic field integrals, the gradient of $\phi$ of
eq.~\eqref{eq:phi-b0}, across the wave function domain (the grid in the
simulation for highest separation $\Delta Z = 40\delta$ consists of 800
million points) was too time-consuming.
Since the whole simulation was performed on a distributed computing cluster;
we divided the domain into subdomains, with each subdomain being simulated on
a separate processor.
In each subdomain, then, the magnetic field $\vec{B}$ was approximated by a
second-order Taylor approximation, centered at the weighted position of the
wave function chunk in a particular domain, and there was no loss of accuracy
from this.  

The overall computational resources spent in the simulation of $\Delta Z = 40\delta$ were as follows. The simulation was run on 40 cores in parallel, with a special parallelized FFT routine as the only step requiring inter-core communication. 
The computation time spent per core was 15 hours, a total of 600 hours. 
The breakdown of the average time spent in a single step of the time evolution loop: 30\% spent on performing FFTs, 39\% on potential propagators (costly because of field evaluations discussed in the preceding paragraph), 28\% on observables calculations, and 3\% on momentum propagators. The overall memory allocation per core was 3GB, a total of 120GB.

\end{paracol}
\reftitle{References}

\end{document}